\documentclass[conference]{IEEEtran}
\IEEEoverridecommandlockouts
\usepackage{cite}
\usepackage{amsmath,amssymb,amsfonts}
\usepackage{algorithmic}
\usepackage{graphicx}
\usepackage{textcomp}
\usepackage{xcolor}
\usepackage{wrapfig}
\usepackage{textcomp}
\usepackage{xcolor}
\usepackage{soul}
\usepackage{braket}
\usepackage{algorithmic}
\usepackage[ruled,vlined, linesnumbered]{algorithm2e}
\usepackage{graphicx}
\usepackage{cite}
\usepackage{amsmath,amssymb,amsfonts}
\usepackage{algorithmic}
\usepackage{graphicx}
\usepackage{textcomp}
\usepackage{xcolor}
\usepackage{wrapfig}
\usepackage{textcomp}
\usepackage{xcolor}
\usepackage{soul}
\usepackage{braket}
\usepackage{algorithmic}
\usepackage[ruled,vlined, linesnumbered]{algorithm2e}
\usepackage{graphicx}
\usepackage{cite}
\usepackage{amsmath,amssymb,amsfonts}
\usepackage{algorithmic}
\usepackage{graphicx}
\usepackage{textcomp}
\usepackage{xcolor}
\def\BibTeX{{\rm B\kern-.05em{\sc i\kern-.025em b}\kern-.08em
    T\kern-.1667em\lower.7ex\hbox{E}\kern-.125emX}}
\begin{document}

\title{Quantum Quandaries: Unraveling Encoding Vulnerabilities in Quantum Neural Networks}









\author{\IEEEauthorblockN{Suryansh Upadhyay}
\textit{The Pennsylvania State University}\\
PA,USA \\
sju5079@psu.edu
\and


\and

\IEEEauthorblockN{Swaroop Ghosh}
\textit{The Pennsylvania State University}\\
PA,USA \\
szg212@psu.edu
}





\maketitle







\begin{abstract}

Quantum computing (QC) has the potential to revolutionize fields like machine learning, security, and healthcare. Quantum machine learning (QML) has emerged as a promising area, enhancing learning algorithms using quantum computers. However, QML models are lucrative targets due to their high training costs and extensive training times. The scarcity of quantum resources and long wait times further exacerbate the challenge. Additionally, QML providers may rely on third-party quantum clouds for hosting models, exposing them and their training data to potential threats. As QML-as-a-Service (QMLaaS) becomes more prevalent, reliance on third-party quantum clouds poses a significant security risk. This work demonstrates that adversaries in quantum cloud environments can exploit white-box access to QML models to infer the user’s encoding scheme by analyzing circuit transpilation artifacts. The extracted data can be reused for training clone models or sold for profit. We validate the proposed attack through simulations, achieving high accuracy in distinguishing between encoding schemes. We report that $\approx$95\% of the time, the encoding can be predicted correctly. To mitigate this threat, we propose a transient obfuscation layer that masks encoding fingerprints using randomized rotations and entanglement, reducing adversarial detection to near-random chance $\approx$42\%, with a depth overhead of $\approx$8.5\% for a 5-layer QNN design.

\end{abstract}





\maketitle

\section{Introduction}

Quantum computing (QC) is gaining attention for its potential to revolutionize problem-solving across various fields. By utilizing quantum properties like superposition, entanglement, and interference, QC offers significant speedups for certain tasks, surpassing classical computing capabilities. With potential applications in machine learning \cite{cong2019quantum,biamonte2017quantum}, security \cite{ghosh2023primer}, drug discovery \cite{cao2018potential}, optimization \cite{farhi2014quantum,tilly2022variational}, finance \cite{orus2019quantum}, and healthcare \cite{gupta2023quantum}, quantum computing is becoming increasingly important in both academia and industry.

Noisy Intermediate-Scale Quantum (NISQ) devices, characterized by a limited number of qubits and susceptibility to noise, face challenges such as restricted qubit connectivity, gate errors, and decoherence, leading to inaccuracies in computation. These limitations hinder the direct implementation of large-scale quantum algorithms, prompting the exploration of hybrid approaches like the Quantum Approximate Optimization Algorithm (QAOA) and Variational Quantum Eigensolver (VQE), which combine classical optimization with quantum subroutines to mitigate noise effects. In this emergent field, quantum machine learning (QML) has gained considerable attention, aiming to improve learning algorithms by leveraging quantum capabilities. Various QML models have been explored, including quantum support vector machines (QSVMs) \cite{rebentrost2014quantum}, quantum convolutional neural networks (QCNNs) \cite{cong2019quantum}, and quantum generative adversarial networks (QGANs) \cite{lloyd2018quantum}. Among these, quantum neural networks (QNNs) \cite{abbas2021power} are notable for replicating the structure and function of classical neural networks within a quantum framework. QNNs utilize parameterized quantum circuits (PQCs) composed of trainable single-qubit and two-qubit gates, with their parameters optimized using classical optimizers. By encoding data into quantum states through methods such as amplitude, angle, or basis encoding, QNNs leverage superposition and entanglement to process information in ways that classical systems cannot replicate. However, optimizing PQCs poses significant challenges due to high computational complexity, resource constraints, and lengthy execution times.

\begin{figure*}
    \centering
    \includegraphics[width= 7in]{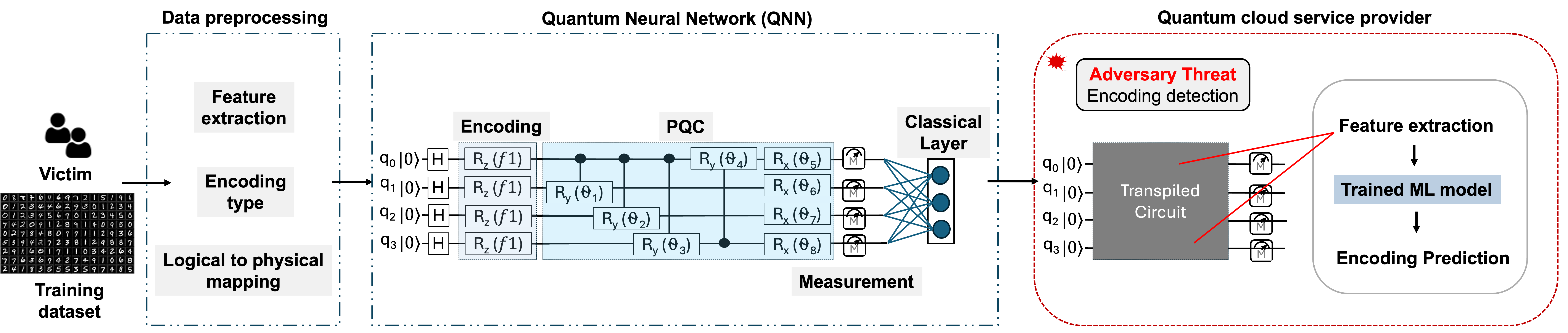}
    \caption{Proposed attack model where the adversary, posing as a reliable quantum cloud service provider, uses the white box access to the QNN submitted to a quantum cloud to identify the type of encoding.}
    \label{threat model}
\end{figure*}

\textbf{Motivation:}
The practical realization of this transformative potential of quantum neural networks (QNNs) largely depends on cloud-based quantum services, where users submit circuits to remote hardware—a dependency that introduces critical, yet understudied, security vulnerabilities. Leading platforms such as IBM \cite{IBMQuantum}, Google \cite{GoogleQuantumComputer}, and AWS Braket \cite{AmazonBraket} provide scalable and accessible quantum computing services; however, they face challenges like job submission latency, queue congestion, and high operational costs. The reliance on cloud infrastructure is driven by the exorbitant costs and specialized requirements of quantum hardware. Training QML models incurs significant expenses, with leading platforms like IBM and IonQ charging up to \$1.6 per second for superconducting qubits and \$0.01 per shot for trapped-ion systems—orders of magnitude costlier than classical alternatives. Compounding this, the iterative nature of hybrid quantum-classical algorithms leads to prolonged training times, often spanning months due to scarce quantum resources and queue congestion on cloud platforms. These factors render trained QML models exceptionally valuable intellectual property (IP), incentivizing adversarial attacks aimed at stealing circuit architectures, optimized parameters, or embedded training data. As the quantum ecosystem expands, third-party providers offering ``Quantum Machine Learning as a Service" (QMLaaS) \cite{MLaaS} further exacerbate these risks. Services like Orquestra \cite{Orquestra} and tKet \cite{pytket} facilitate multi-hardware integration, while Baidu's ``Liang Xi" \cite{baidu} offers flexible quantum access via mobile, desktop, and cloud interfaces. While trusted hardware remains the preferred choice for applications with high economic or social stakes, hybrid quantum-classical algorithms often incur significant expenses and delays due to the numerous iterations required. Although governments and large enterprises may possess dedicated quantum resources, they are costly and geographically constrained. This growing reliance on third-party compilers, hardware ecosystems, and cloud services introduces pressing concerns regarding the reliability and security of quantum computations.

A QNN comprises of a data encoding circuit that transforms data into quantum states, a parameterized quantum circuit (PQC) with tunable parameters, and measurement operations for extracting information. Encoding methods, such as basis, amplitude, or angle encoding, are crucial as they determine how input data is represented in the quantum system. These methods imprint distinct structural signatures in transpiled circuits, such as characteristic gate sequences, rotation angle distributions, and entanglement patterns. Malicious cloud providers with white-box access, acting as adversaries, can exploit these artifacts to reverse-engineer proprietary encoding schemes, enabling them to infer sensitive details about the input data, model architecture, and even training objectives. Such breaches compromise not only intellectual property but also data privacy, as encoding methods often embed domain-specific preprocessing critical to a model’s functionality. Several studies have addressed protecting quantum circuits from untrusted clouds \cite{upadhyay2022robust, upadhyay2023trustworthy}, however they focus on protecting computation outputs or mitigating hardware-level attacks, overlooking the risks posed by encoding side channels. This gap leaves users vulnerable to adversaries who can simply inspect transpiled circuits and steal proprietary encoding techniques or reconstruct training data. \textit{This paper to our knowledge is the very first attempt which aims to understand and mitigate this security vulnerability}. 

\textbf{Contributions:} In this work, we:
(a) propose a novel framework for classifying popular quantum encoding schemes (i.e., angle, amplitude, and basis encodings) through transpilation patterns, (b) demonstrate the effectiveness of our approach using a comprehensive dataset of circuits generated with varied encoding types and parameterized quantum circuits (PQCs), (c) introduce a multi-level feature extraction pipeline that captures discriminative patterns from transpiled circuits, including structural signatures, rotation analysis, and entanglement metrics, (d) validate our methods through simulations, achieving high accuracy in distinguishing between encoding schemes, (e) propose a lightweight, transient obfuscation layer that masks encoding fingerprints using randomized rotations and entanglement, reducing adversarial detection to near-random chance.

\textbf{Paper organization:} Section II provides background information. Section III outlines the threat model and section IV presents the proposed defense. Section V covers results and Section VI concludes the paper.

\section{Background}



\subsection{Quantum Neural Network (QNN)}

A QNN consists of three main components (Fig. \ref{threat model}): (i) a data encoding circuit that maps classical data into quantum states, (ii) a parameterized quantum circuit (PQC) with tunable parameters, and (iii) measurement operations to extract useful information from the quantum system. Due to the limited number of qubits in current quantum devices, classical preprocessing techniques such as Principal Component Analysis (PCA) are often employed to reduce the dimensionality of input data. The reduced data is then encoded into quantum states using methods like basis encoding, amplitude encoding, or angle encoding. The PQC is the core trainable component of a QNN. It consists of a sequence of quantum gates with adjustable parameters designed to recognize patterns in data and solve specific problems. Finally, measurement operations collapse the qubit states to either $0$ or $1$. The expectation value of Pauli-Z is used to determine the average state of the qubits. These measured values are typically fed into a classical neuron layer, with the number of neurons equal to the number of classes in the dataset. This layer performs the final classification task. A classical optimizer optimizes the parameters iteratively to achieve the desired input-output relationship.

\subsection{Quantum Cloud Services}

Quantum computers require extensive and expensive infrastructure, such as cryogenic coolers and superconducting wires, making direct access challenging. Quantum cloud services like IBM \cite{IBMQuantum}, Google \cite{GoogleQuantumComputer}, and AWS Braket \cite{AmazonBraket} provide remote access, simplifying system management. Users send their quantum circuits to these services, specifying target hardware and metadata. 

\subsection{Data Encoding}

Data encoding is a critical step in quantum machine learning (QML) and quantum neural networks (QNNs), as it transforms classical data into quantum states, enabling quantum algorithms to process information efficiently. The choice of encoding method significantly impacts the performance, expressivity, and scalability of quantum models. Some widely used encoding schemes are:

\subsubsection{Basis Encoding}  

 Maps classical data directly to the computational basis states of qubits. In this approach, each classical data point is represented as a binary string, where each bit in the string corresponds to a qubit state. For example, consider a classical data vector \( [1, 0, 1] \), which can be encoded into the quantum state \( \ket{101} \). This encoding method is particularly advantageous for representing discrete or categorical data, as it provides a straightforward and efficient mapping from classical to quantum representation. However, a major drawback is that the number of qubits required grows linearly with the size of the dataset, which poses scalability challenges when dealing with high-dimensional data. 

 \subsubsection{Angle Encoding}  

Embeds classical data into the rotational angles of qubits on the Bloch sphere. In this approach, a classical feature \( x_i \) is represented as a rotation along a specific axis using quantum gates such as \( R_x(x_i) \), \( R_y(x_i) \), or \( R_z(x_i) \). This method allows for the encoding of continuous data into quantum states efficiently. Additionally, dense angle encoding can encode multiple features per qubit by incorporating additional phase gates, enabling more compact representations. A key advantage of angle encoding is its efficiency in terms of gate depth and the reduced number of qubits required compared to basis encoding, making it particularly suitable for continuous-valued data.

\subsubsection{Amplitude Encoding}  

Utilizes the amplitudes of quantum states to represent normalized classical data. Given a classical vector \( x = [x_1, x_2, \dots] \), it is transformed into the quantum state \( \ket{\psi} = \sum_{i} x_i \ket{i} \), where each amplitude \( x_i \) corresponds to a feature in the data. This encoding scheme is highly space-efficient, as it can represent \( 2^n \) features using only \( n \) qubits, making it particularly advantageous for large datasets. Amplitude encoding is primarily limited by its high circuit complexity, requiring intricate quantum operations for precise state preparation.

\subsubsection{Hybrid and Advanced Encoding Techniques}  

Recent research has explored hybrid approaches that combine angle and amplitude encoding to leverage their respective strengths. These hybrid encoding methods aim to enhance information representation while addressing the limitations inherent to individual techniques.


\subsection{Related works}

The security of machine learning systems has been extensively studied in classical computing, with attacks ranging from side-channel exploits on hardware accelerators to black-box model extraction via adversarial queries \cite{hua2018reverse}\cite{oh2019towards}. Recent work has extended reverse engineering (RE) to quantum circuits, using lookup tables (LUTs) to map transpiled gate sequences to original QML architectures \cite{ghosh2024quantum}. By analyzing rotation gate ordering and entanglement patterns, adversaries can infer circuit parameters, exposing proprietary model designs. Building on this, \cite{upadhyay2024quantumdatabreachreusing} demonstrates how adversaries can extract state preparation circuits and training labels from QML models, directly stealing training data by reverse-engineering the encoding process. This underscores the criticality of encoding schemes as attack surfaces, as they bridge raw data to quantum computation. Quantum homomorphic encryption (QHE) \cite{fisher2014quantum}, while theoretically viable, imposes prohibitive overheads incompatible with near-term devices. Recent quantum-specific defenses address model theft through strategies like distributed execution (QuMoS \cite{wang2023qumos}) and output obfuscation (STIQ \cite{kundu2024stiq}), but these focus on protecting trained parameters rather than preventing encoding detection. Furthermore, while strategies like circuit partitioning \cite{upadhyay2022robust,upadhyay2023trustworthy} aim to distribute trust across providers, they fail to protect QNNs and related IPs because any untrusted provider with access to transpiled circuits can recover encoding schemes. Our work diverges by addressing encoding-specific transpilation artifacts. Unlike \cite{upadhyay2024quantumdatabreachreusing}, which focuses on training data extraction, we demonstrate that adversaries can preemptively identify encoding methods to streamline subsequent attacks.

\section{Threat model}

\subsection{Basic Idea}

The assets in QNNs include proprietary algorithms, training data, and the resulting trained models. Identifying the encoding scheme used in a QNN can facilitate the extraction of embedded features, posing a security risk. While quantum encoding methods such as angle, amplitude, and basis encoding provide distinct data representations, their implementation on near-term hardware introduces artifacts due to transpilation. We propose that these artifacts inherently imprint encoding-specific signatures during transpilation, which can be exploited by a malicious quantum service provider. In the proposed attack model, the adversary operates as an untrusted cloud provider while posing as a legitimate and reliable hardware vendor. With access to transpiled QNNs submitted for training, the adversary can analyze the state preparation (encoding) circuits (Fig. \ref{threat model}). The primary objective is to extract critical details about the victim’s encoding scheme and embedded features, which can then be monetized.

\subsection{Adversary Capabilities and Assumption}

We assume that: (a) The adversary has access to the transpiled circuit submitted by users and the results produced by the quantum computer. This is justified because the cloud provider, by design, has access to both the QNN and the measurement results; (b) The victim employs one of the three widely used encoding schemes: angle encoding, amplitude encoding, or basis encoding, and encodes only one feature per qubit. This assumption is reasonable as these encodings are the foundation of many near-term QML algorithms and are widely adopted due to their compatibility with current NISQ devices. By restricting our scope to these encoding methods, we establish a realistic and practical baseline for analyzing encoding detection under existing hardware constraints.

\subsection{Attack Process}

This section presents a hybrid quantum-classical approach for encoding classification, combining quantum circuit transpilation patterns with classical feature engineering. The methodology focuses on identifying quantum circuit signatures through systematic analysis of gate-level implementations across different encoding schemes.

\subsubsection{\textbf{Data Set Generation}}

The dataset is constructed using PQCs with varied encoding schemes and rotational configurations:

\textbf{Encoding Layer Construction}: We generate distinct quantum state representations using three fundamental state preparations: \textit{a) Pauli Rotations}: maps classical features $x_i$ to qubit rotations $R_j(x_i)$, where $j \in {X,Y,Z}, $with $x_i \sim \mathcal{U}(0, 2\pi)$. \textit{b) Amplitude Encoding}: represents $2^n$ features in $n$-qubit state using haar-random $2^n$-dimensional statevector initialization $\|\psi\rangle = \frac{1}{\|\vec{v}\|}\sum_k v_k|k\rangle$. \textit{c) Basis Encoding}: direct bit-to-qubit mapping via X gates.
    
\textbf{PQC Augmentation}: PQCs with varying configurations (e.g., entanglement patterns and gate types) are added to enhance the diversity of the dataset. We used different PQCs with different starting gates (P):
    \begin{equation*}
        \mathcal{P} = \{\texttt{rx}, \texttt{ry}, \texttt{rz}, \texttt{cx}, \texttt{x}, \texttt{sx}, \texttt{crx}, \texttt{cry}, \texttt{crz}\}
    \end{equation*}
    
\textbf{Hardware Simulation}: Quantum circuits are transpiled using a noisy fake back-end (GenericBackendV2), ensuring features reflect hardware-aware implementations.

\textbf{Labeling}: Each sample is labeled based on its encoding type to facilitate supervised learning.

\textbf{Example:} For a sample 3-qubit system with single-feature encoding, we generate a comprehensive dataset of 18,000 circuits.

\begin{figure}
    \centering
    \includegraphics[width= 3.5in]{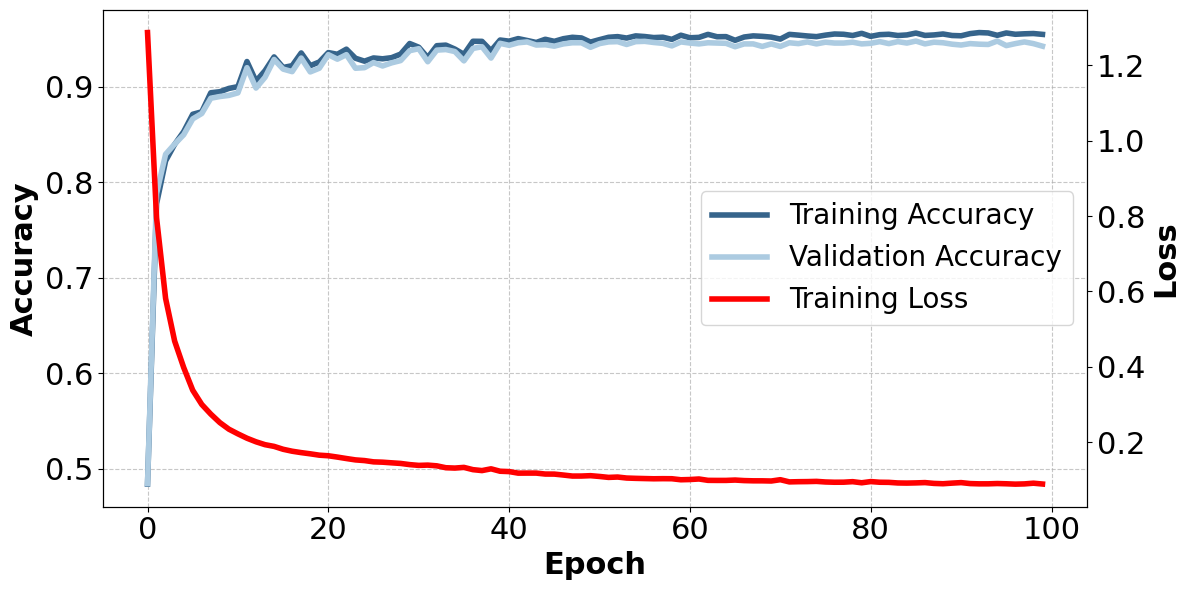}
    \caption{ Training, validation accuracy and loss for classifying the encoding scheme in a 3-qubit circuit with 3 encoded features.}
    \label{train}
\end{figure}

\subsubsection{\textbf{Feature Extraction and Engineering}}
Feature extraction focuses on converting the characteristics of quantum circuits into a numerical feature space suitable for classical machine learning. Our approach starts with an intuitive analysis of transpiled circuits across various encoding schemes to identify recognizable patterns. Insights from this exploratory analysis guided the creation of a systematic feature extraction process, designed to capture key quantum circuit patterns crucial for differentiating between various encoding types.

\begin{equation}
\text{Features} = 27 + 2 \times \text{Qubits}
\end{equation}

The feature engineering pipeline refines the process by identifying discriminative circuit features (Equation 1) through three levels of analysis:

\textbf{Structural Signatures:} Provide essential insights into the encoding mechanisms used within quantum circuits. Key markers include: \textit{a) Basis encoding markers}: Ratios of X and SX gates, along with binary pre-CNOT patterns; \textit{b) Amplitude encoding indicators}: Diversity in rotation gate parameters indicating high variability in amplitude encoding; \textit{c) Transpilation artifacts}: Frequency of RZ-SX sequences and statistical distributions of rotation angles.

\textbf{Rotation Analysis:} Rotation-based features capture the statistical properties of gate parameters, including: a) Mean, standard deviation, and entropy of rotation angles for each gate type; b) Correlation of consecutive gate parameters across the circuit; c) Modular distributions of rotation angles $(\theta \mod 2\pi)$ to identify characteristic transformations.

\textbf{Entanglement Signatures:} Entanglement metrics characterize multi-qubit interactions, which are crucial for understanding circuit complexity.

The extracted feature space captures both global circuit properties and local qubit-wise patterns.

\subsubsection{\textbf{Training Model}}

The classification model integrates quantum-aware preprocessing with classical deep learning to effectively differentiate encoding schemes.

\textbf{Preprocessing Pipeline:} The preprocessing pipeline ensures standardized input representation and maintains encoding balance. A stratified data splitting approach is employed, using a 60-20-20 split to preserve encoding ratios across training, validation, and test sets. Additionally, Z-score normalization is applied to standardize the features to have zero mean and unit variance.

\textbf{Network Architecture:} The neural network employs a multilayer perceptron (MLP) classifier designed to capture both local and global circuit characteristics. The input layer consists of a number of neurons equal to the total extracted features, ensuring a direct mapping from the feature space to the network. The first hidden layer, comprising 25 neurons, while the second hidden layer, with 10 neurons. Finally, the output layer applies a softmax activation function to classify inputs into five encoding types. The network is trained using the ReLU activation function, the Adam optimizer, and a maximum of 100 training iterations.

\begin{table}[!t]
\caption{Classification Performance Metrics (3-qubit QNN)}
\label{table:performance}
\centering
\begin{tabular}{lccccc}
\hline
\textbf{Encoding Type} & &\textbf{Precision} & \textbf{Recall} & \textbf{F1-Score} & \textbf{Support} \\
\hline
Amplitude & &1.00 & 1.00 & 1.00 & 900 \\
\hline
Basis & & 0.93 & 0.81 & 0.87 & 900 \\
\hline
&Rx & 0.92 & 0.96 & 0.94 & 900 \\
Angle &Ry & 0.93 & 0.97 & 0.95 & 900 \\
&Rz & 0.95 & 0.98 & 0.96 & 900 \\
\hline
Accuracy & &- & - & 0.94 & 4500 \\
Macro Avg & & 0.94 & 0.94 & 0.94 & 4500 \\
Weighted Avg & &0.94 & 0.94 & 0.94 & 4500 \\
\hline
\end{tabular}
\end{table}

\section{Proposed Defense}
The adversarial detection of quantum encoding scheme relies on identifying predictable patterns in the transpiled circuit structure. To counter this, we introduce a temporary scrambling mechanism that strategically alters the transpiled circuit’s transient properties while preserving its final output. Our proposed defense mechanism acts as a transient ``cloaking" layer strategically positioned between the quantum encoding circuit and the parameterized quantum circuit (PQC). This layer temporarily obfuscates the encoded state’s structure, ensuring adversaries cannot identify the encoding methods. The defense operates in three stages:

\subsubsection{\textbf{Obfuscation Phase}} Immediately after the encoding circuit, we apply a sequence of randomized quantum gates:

\textit{Basis randomization:} Hadamard (H) gates place all qubits into superposition states.

\textit{Phase randomization:} Parameterized RX($\theta$) rotations with angles $\theta$ sampled uniformly from [-$\pi$, $\pi$] introduce qubit-specific phase shifts.

\textit{Entanglement creation:} CNOT gates entangle adjacent qubit pairs (e.g., qubit 0$\rightarrow$1, 2$\rightarrow$3), generating artificial correlations.

After the initial state encoding , we apply:
\begin{equation*}
U_{\text{obf}} = \left(\prod_{i=0}^{n-1} H_i RX(\theta_i)\right) \left(\prod_{j=0}^{\lfloor n/2 \rfloor} \text{CNOT}_{2j,2j+1}\right)
\end{equation*}

\begin{figure}
    \centering
    \includegraphics[width= 3.5in]{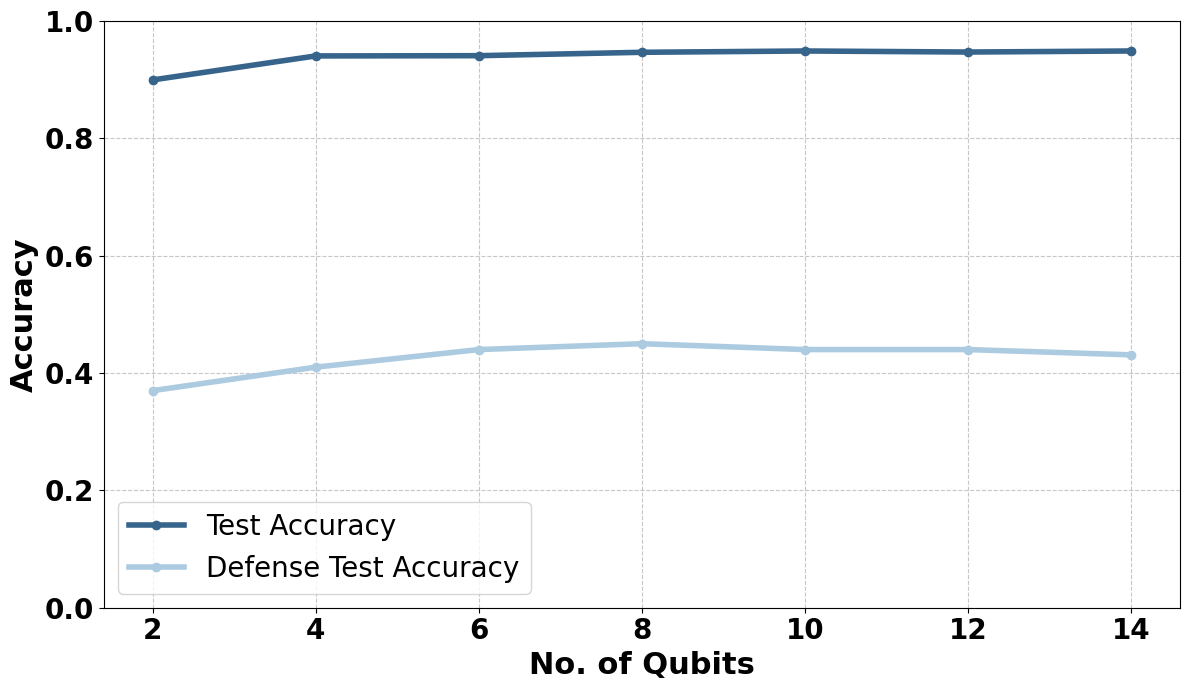}
    \caption{Test accuracy for circuits with varying qubit counts, where each qubit encodes a single feature.}
    \label{var_qubits}
\end{figure}

\subsubsection{\textbf{Isolation Barrier}} A hardware-enforced barrier prevents quantum compilers from optimizing or rearranging gates across the defense structure, preserving the intentional obfuscation pattern. The barrier is not observable at the cloud end during execution thereby eliminating any adversarial clue. 

\subsubsection{\textbf{Inversion Phase}} Before the PQC circuit, we systematically undo the obfuscation:

\textit{Entanglement removal:} Apply inverse CNOT gates in reverse order (e.g., qubit 2$\rightarrow$3 first, then 0$\rightarrow$1).

\textit{Phase cancellation:} Implement RX($\theta$) rotations to negate the initial random phases.

\textit{Basis restoration:} Reapply H gates to return qubits to their original basis.

This architecture preserves encoding anonymity while ensuring correct functionality, as the parameterized quantum circuit (PQC) operates directly on the originally encoded quantum state \( \ket{\psi_{\text{enc}}} \).

\section{Results and evaluation}

Due to the long queue times and limited availability of real quantum devices, we used Qiskit's fake provider module that mimics IBM's system and includes real hardware-calibrated data. Training was performed on an Intel Core-i7-12700H CPU with 40GB of RAM. We evaluated the efficacy of the proposed encoding detection attack through three critical dimensions: classification accuracy across QNN architectures with varying qubit counts and different PQCs, feature scalability, and training convergence.

\subsection{Training Convergence and Attack Reliability}

The training dynamics Fig.\ref{train} confirms the reliability of the classifier under the design feature space. The validation accuracy reaches 80\% in epoch 40 and stabilizes at 90\% training accuracy by epoch 100. Concurrently, training loss decreases monotonically from 1.2 to 0.2, indicating stable optimization without overfitting.

Table I details the attack’s precision, recall, and F1-scores for distinguishing encoding schemes in a 3-qubit system. The amplitude encoding achieves perfect classification (F1=1.00), attributed to its unique transpilation artifacts. In contrast, basis encoding shows reduced recall (0.81) and F1-score (0.87), likely due to its reliance on X and SX gates—patterns that may overlap with angle-encoded circuits post-transpilation. The adversary can predict the encoding scheme used by the user with a $\approx$ 95\% success rate. \textit{These results validate the feature engineering pipeline’s ability to isolate encoding-specific signatures, even for structurally similar angle variants. The adversary can identify the type of encoding used and, in the case of angle encoding, determine the specific rotation gate employed for data encoding}.

\begin{table}[!t]
\caption{Average Circuit Depth (Original vs Obfuscated)}
\label{table:depth}
\centering
\begin{tabular}{lccccc}
\hline
\textbf{Encoding Type} && \textbf{Original} & \textbf{Obfuscated} & \textbf{$\Delta$ Abs} & \textbf{$\Delta$$\uparrow$\%} \\
\hline
Amplitude && 95.2 & 101.1 & +5.9 & 6.2\% \\
\hline
Basis && 67.6 & 74.1 & +6.5 & 9.7\% \\
\hline
&Rx & 70.2 & 76.1 & +5.9 & 8.4\% \\
Angle &Ry & 69.2 & 76.1 & +6.9 & 10\% \\
&Rz & 67.7 & 74.1 & +6.4 & 9.5\% \\

\hline
\end{tabular}
\end{table}

\subsection{Feature Scalability and Attack Robustness}

The linear relationship between feature count and the number of qubits (Equation.1) demonstrates the attack’s adaptability to circuit complexity. At the 3-qubit baseline, 33 features capture structural patterns, rotation statistics, and entanglement properties. Expanding to 14 qubits increases the feature space to 55, improving the ability to distinguish encoding-specific transpilation artifacts (e.g., RZ-SX sequences in basis encoding). While extrapolation to 100 qubits (227 features) raises classical computational costs, \textit{the linear growth ensures practical viability for near-term attacks targeting $\leq 14-qubit$ QNNs—consistent with current QML benchmarks.}

\subsection{Encoding Classification Accuracy vs. Qubit Count}
The adversary can accurately classify encoding types even as the number of qubits increases (Fig.\ref{var_qubits}). For a 3-qubit system, the attack achieves 94\% accuracy, demonstrating its effectiveness at small scales. As qubit count scales to 14, test accuracy improves from 90\% to 95\%, indicating that the model effectively captures and differentiates encoding-specific transpilation artifacts. Furthermore, the model not only identifies the encoding type but also determines the specific rotation gates used in angle encoding, even as circuit complexity grows. \textit{The transpilation artifacts and circuit structures remain sufficiently distinct at higher qubit counts, allowing the model to generalize effectively across varying circuit sizes.}

\subsection{Defense Efficacy and Overhead}

To assess the effectiveness of the proposed defense, we generated a test dataset incorporating the defense strategy and evaluated the trained classifier's accuracy. The results indicate a significant reduction in adversarial encoding detection accuracy, dropping from 95\% to an average of 42\% across all encoding types (Fig.\ref{var_qubits}). However, this obfuscation introduces an average depth increase of $\approx$8.5\% compared to baseline transpiled circuits. Table II presents a comparison of the average circuit depth across 800 instances for each encoding type, considering a 5-layers, of a low-depth PQC. Notably, for deeper QNNs and multi-layer PQC architectures, the relative \% increase in depth is expected to be negligible.


\section{Conclusion}

This work identifies a critical vulnerability of QNN’s white-box access to adversaries in untrusted quantum cloud by demonstrating that quantum encoding schemes can be reliably detected with $\approx$95\% accuracy. This is due to transpilation artifacts which can aid in subsequent state preparation circuit (an IP) theft. To mitigate this risk, we propose to strategically insert transient obfuscation layers—randomized rotations and entanglement—to obscure encoding patterns. This approach reduces adversarial detection accuracy to near-random levels ($\approx$42\%) while introducing a minimal circuit depth overhead of $\approx$8.5\% for a 5-layer QNN design.

\section{Acknowledgment}

This work is supported in parts by NSF ( CNS-2129675, CCF-2210963, CCF-1718474, and DGE-2113839) and Intel's gift.

\bibliographystyle{IEEEtran}
\bibliography{asp}
\end{document}